# Enhanced Bloom's Educational Taxonomy for Fostering Information Literacy in the Era of Large Language Models

Yiming Luo [1], Ting Liu [1], Patrick Cheong-Iao Pang [1*], Dana McKay [2], Ziqi Chen [1], George Buchanan [2], Shanton Chang [3]

[1] Faculty of Applied Sciences, Macao Polytechnic University

[2] School of Computing Technologies, RMIT University

[3] School of Computing and Information Systems, The University of Melbourne

## Abstract

The advent of Large Language Models (LLMs) has profoundly transformed the paradigms of information retrieval and problem-solving, enabling students to access information acquisition more efficiently to support learning. However, there is currently a lack of standardized evaluation frameworks that guide learners in effectively leveraging LLMs. This paper proposes an LLM-driven Bloom's Educational Taxonomy that aims to recognize and evaluate students' information literacy (IL) with LLMs, and to formalize and guide students practice-based activities of using LLMs to solve complex problems. The framework delineates the IL corresponding to the cognitive abilities required to use LLM into two distinct stages: Exploration & Action and Creation & Metacognition. It further subdivides these into seven phases: Perceiving, Searching, Reasoning, Interacting, Evaluating, Organizing, and Curating. Through the case presentation, the analysis demonstrates the framework's applicability and feasibility, supporting its role in fostering IL among students with varying levels of prior knowledge. This framework fills the existing gap in the analysis of LLM usage frameworks and provides theoretical support for guiding learners to improve IL.

## 1. Introduction

LLMs are state-of-the-art artificial intelligence (AI) systems designed to process and generate information, trained in a self-supervised environment on extensive corpora, facilitating significant advancements in performance in translation and summarization [1]. The integration of LLM-based technology has been proven to assist students to learning by utilizing ChatGPT to retrieve information [2] or summarize text more efficiently [3]. With the continuous development of AI, higher education students now have access to abundant learning resources easily through LLMs [4]. However, the challenge posed by information overload is how students can find, evaluate and effectively use this information [5].

Arguably information literacy (IL) has long been seen as one of the core human skills in modern life [6], but it has grown in importance over the past few years due to the emergence of misinformation [7]. Information retrieval (IR) ability is one of the basic parts of IL, which is initially defined as the process of recognizing information needs, acquiring information sources, evaluating the quality of information, and selecting the appropriate information [8]. Using information to solve practical problems, commonly referred to as "problem-solving skills", involves the ability to organically evaluate, organize, and integrate information, and apply this information to specific contexts to find suitable application solutions [9].

IL has proven to have a positive impact on students' success today by promoting students to engage in integrative learning [10] and critical thinking [11]. Despite increasing empirical evidence that indicates significant potential of LLMs to assist higher educational students in automating and customizing a variety of academic tasks including professional research [12], there remains a lack of practice recommendations and regulatory guidelines for their integration into student problem-solving based on IR [13]. Compared to the rapid growth of LLMs in the field of educational technology, research on their evaluation and regulation within teaching has been relatively slow. Most studies focus on outcomes such as learning achievements [14][15], while a growing body of research emphasizes that teacher-assisted LLM technology can benefit students [16] [17]. Teachers in higher education need to balance teaching and research, and the energy available to guide students in practical situations is limited [18].

Bloom's Educational Taxonomy (BET), as a classic classification method that categorizes educational objectives into different levels based on the complexity of human cognitive activities, provides theoretical support for learning goal identification [19] and teaching assessment [20]. After multiple revisions and developments by researchers such as Anderson and Krathwohl, Bloom's Taxonomy has evolved to better align with the needs of contemporary educational stakeholders [21]. IL is critical for students' lifelong learning development [22]; however, there is currently a lack of research assessing students' IL skills using Bloom's Taxonomy with the emergence of LLMs. Against this background, this paper proposes the following research questions:

**RQ1**: *How can LLMs adapt to student learning within the framework of BET?*

**RQ2**: *How can the framework guide and assess students' use of LLMS to improve their IL practically?*

To answer these questions, this study introduces an LLM-driven Bloom Educational Taxonomy (LBET), which aims to identify, standardize and guide higher education students to use LLMs to solve professional issues, such as data analysis, in environments with limited teacher supervision. LBET divides higher education students' IL level of using LLMs into two major stages and seven levels. Through a case analysis, the study also demonstrates how the LBET framework evaluates students' IL levels when using LLMs and preliminarily verifies that students can utilize this framework to enhance their abilities in IR and complex task processing, providing a comprehensive guide for higher education to foster students' IL in the era of LLMs.

## 2. Dilemma in the application of LLMs in education

### 2.1 Difficulties faced by students using LLMs

LLMs provide great convenience in the learning process, including time efficiency, personalized homework feedback, and writing assistance [23][24]. While LLMs can simplify the retrieval of relevant information, students may overestimate their accuracy or depth, leading to an undue sense of trust and dependency [25].

There is some evidence pointing to potential cognitive biases in using information [26], and challenges in applying appropriate LLMs strategies in complex situations [27]. Research by Kazemitabaar et al. [28] shows that learners who combine their own coding strategies with AI suggestions perform better than those relying solely on AI-provided solutions. Stadler indicated that although LLMs reduce mental work, they may harm the depth of students' scientific inquiry [29]. Given that LLMs typically do not provide transparent source citations, students may struggle to verify the accuracy and reliability of the information provided. This is particularly concerning in academic settings, where plagiarism, incorrect citations, and improper language use can lead to serious consequences [30]. These studies highlight the potential risk of over-reliance on LLMs in a limited learning environment with no teacher involvement. Without the necessary review frameworks and analytical skills, students may become accustomed to accepting unverified information, forming a cognitive information cocoon that undermines their long-term IL development [31]. Therefore, a comprehensive framework for assessing students' use of LLMs is urgently needed.

**2.2 The challenges of guiding and regulating LLMs**

Existing guidelines have discussed ethical issues regarding the use of LLMs, including safety and morality [32], Floridi [33] has highlighted the importance of separating intelligence and agency in LLMs, stressing the need for cautious design and supervision to ensure proper control of these systems. A classic example of moral hazard in practice is the problem of bias inherent in the output of LLMs. Without proper guidance, these biases can be perpetuated or even amplified when students use LLMs uncritically [34]. This may hinder students' ability to independently develop mastery of LLMs.

Other experts have raised the question of whether LLMs help or hinder learning in information seeking [35]. Arguably synthesis of different resources is part of the learning process, and a higher order learning skill [36]. While current standards endeavor to address specific operations to improve automating synthesis performance of LLMs such as prompt engineering [37] or program development [38], these models may still be synthesized in a way that reflects bias (as noted above). That includes misinformation [39], or that does not reflect the most useful synthesis for the student's own learning journey. Understanding the limitations of LLMs, and the impact they may have on their own learning is a key LLM literacy skill for students.

Despite the lack of meticulous supervision and detailed discussion on the application of LLMs, it is undeniable that LLMs are increasingly seen as societal actors with transformative potential, especially in "understanding" complex social systems [40] [41]. LLMs are promoting the transformation of the learning system. This transformation is especially significant in terms of students' exploratory and independent learning processes, particularly under conditions where teacher guidance is limited [42]. These forward-looking applications include intelligent question-answering robots based on LLMs [43], graphical reasoning tools [44], interactive digital learning books [45]. In the field of education, LLMs may participate in learning and teaching activities in a more equal and constructive role in the future, rather than just existing as a simple tool [46].

**3. Bloom's Taxonomy and derivatives**

BET presented a widely recognized system for classifying educational objectives, encompassing three major domains: cognitive, affective, and psychomotor [47]. The cognitive domain was further subdivided hierarchically into the domains of knowledge, comprehension, application, analysis, synthesis, and evaluation. The BET framework is a cumulative, bottom-up hierarchy that forms a continuum from lower order thinking skills (LOTS) to higher order thinking skills (HOTS). In particular, Knowledge, Comprehension, and Application are considered LOTS, and the levels of analysis, synthesis and evaluation are considered to be the embodiment of HOTS.

Knowledge is typically understood as the basic recall of specific facts or abstract concepts, including identifying or retrieving information. Comprehension consists of grasping the meaning of information, such as classifying information and comparing and contrasting concepts. Application refers to using acquired knowledge in new situations to make decisions or perform procedures. Analysis involves methods such as separating, organizing, inferring, and classifying to break down information into more fundamental components to explore relationships and underlying structures. Synthesis involves integrating and designing new ideas based on prior knowledge and learning. Evaluation involves the assessment and judgment of the outcomes of tasks and is regarded as the highest level of cognitive processing [48].

BET is regarded as a valuable measurement tool capable of defining the precise meanings of broad educational goals within a curriculum [36], and acts as a universal language in education to enhance interaction among various educators, subjects, and grade levels [49]. In real-world use, the BET framework proved to be often less precise [50] as the complexity of cognitive demands can vary depending on specific goals, which may make it difficult in practice to fully cover and address cognitive objectives using BET [51]. In response to these challenges, Krathwohl, one of the original contributors to Bloom's Taxonomy, proposed Revised Bloom's Taxonomy (RBT) [44] shown in *Figure 1*. The revisions aimed to improve three key aspects: terminology, structure, and emphasis [52]. Most notably, Krathwohl replaced the noun-based categories of the original taxonomy with verbs, transforming "Knowledge" into "Remembering"; "Comprehension" into "Understanding", and "Synthesis" into "Creating", it introduced four types of knowledge: factual, conceptual, procedural, and metacognitive, allowing educators to more effectively visualize how learning occurs across both the knowledge and cognitive dimensions [48]. Building on these revisions, Darwazeh [53] proposed further refinements, such as repositioning metacognition as the most complex cognitive level and subdividing the "remembering" and "creating" levels into more detailed sub-categories.

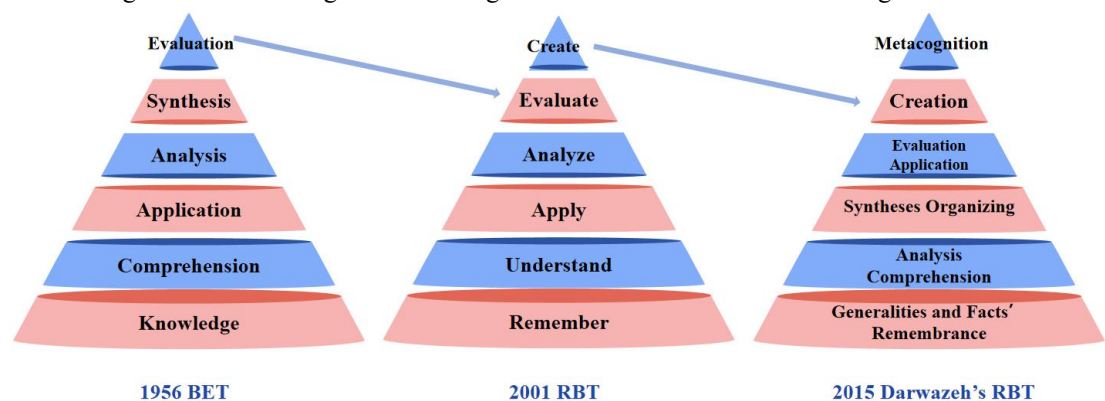

Figure 1. Development of taxonomy theory

## 4. Bloom's Taxonomy in digital learning

Though BET and RBT are fundamental tools for teachers, these frameworks do not address the new problems and objectives arising from technological advancements. [54]. Given the evolving landscape of digital learning resources and opportunities facilitated by educational informatization, Churches first introduced Bloom's Digital Taxonomy (BDT), which is derived from BET, in 2010 to reflect the integration of digital tools in education and guide the learning process under educational informatization [55].

BDT is a key benchmark in modern education's digitalization, serving as a framework for evaluating information and communication technology competencies [56]. As shown in *Table 1*, BDT still retains the pyramid structure of RBT but adapts the content of each level to information technology: e.g. Remembering is explained as recalling, retrieving, and recognizing information, and Applying as the ability to use information in new situations. The corresponding actions for each skill were identified as being consistent with digital competence.

Additionally, both AI and social media tools have made respective innovations within the framework of RBT. Demir [57] proposed a Taxonomy of Social Media for Learning (TSML), which redefined the levels of RBT into three major dimensions: Awareness, Cognition, and Creation, along with five sub-goals. It closely focuses on evaluating social media as a tool for students and teaching. In establishing TSML, like Darwazehs' RBT, Metacognition is added as the highest cognitive level. Metacognition involves not only cognition of cognition but also the process of reflecting upon and adjusting learners' cognitive activities [58]. Metacognition serves as the core of the taxonomy, equipping learners with advanced skills such as critical thinking, practical skills, and global awareness. On the other hand, Faraon introduced Bloom's Taxonomy based on AI (BTA) by defining all levels of AI application and gave some examples of the practical application of basic LLMs tools [59]. To clearly demonstrate the usefulness of the tool, some examples of text generation models represented by ChatGPT [60] and image generation models represented by DALL-E [61] are displayed under the BTA framework in *Table 1*. However, BTA mainly emphasizes the classification of the purpose of AI tools, rather than focusing on users and the learning process itself, implying that the guidance needs and potential risks for students when using these LLMs are overlooked. This paper further expands the classification of these two assessment tools by targeting the design of more comprehensive text-based LLMs in relevant works, with a focus on tracking student behavior to meet the requirements for HOTS in a modern context that emphasizes IL.

Table 1. Bloom digital taxonomy and derived theories

| BDT | TSML (social media) | | BTA (AI) | | |
|---|---|---|---|---|---|
| Level | Description | Example | Description | Example for ChatGPT | Example for DALL-E |
| **P1 - Remembering** Recall and recognize information and concepts | Students **view** course content on private social media sites | Student teachers (ST) viewed course-related resources on group pages | Help Students search for basic knowledge | *ChatGPT* can generate a simple document containing basic concepts and knowledge points to provide students with information | *DALL-E* can generate a clear anatomical diagram of human anatomy to help students remember the names and locations of different bones in a visual way |
| **P2 - Understanding** Understand the meaning of the information | Lecturers **post** knowledge through online content | ST presented their lesson preparation materials within a standard framework | Help students with reading comprehension | *ChatGPT* can generate a concise summary based on the medical paper uploaded by the student to explain its key points | *DALL-E* can generate relevant images about animal anatomy to assist in understanding academic reading in the field of animal care |
| **P3 - Applying** Apply knowledge to new situations | Students **interact** with others on the network by sharing their ideas and insights | ST developed a draft of teaching materials, discussing with their classmates in their groups | Helps students improve their writing performance | *ChatGPT* can output the text in an academic style and help students edit their essays and improve the language of their paragraphs | *DALL-E* can generate images related to the topic to help students better express their views in academic reports |
| **P4 - Analyzing** Break down information to understand structure | Students collect valuable ideas from others and **analyze** them to put them into | ST analyzed the ideas and comments collected from their peers and applied them to their own work | Handles code errors in web development | *ChatGPT* can support learners when they encounter coding errors | *DALL-E* cannot be used directly for code problem, but it can help students break down complex code and |
| | practice | | | | structure into visual parts to better understand how the code works |
| **P5 - Evaluating** Make judgments and provide reasoning | Students **evaluate** the strengths and weaknesses of their work by self-evaluation and peer evaluation | ST uploaded the teaching material and showed them to the group for feedback | Help students use critical thinking when writing scientific reports | *ChatGPT* can check the logic and coherence of the argument and give overall improvement suggestions | *DALL-E* can generate scenarios under different experimental conditions to help students consider the likelihood of success of the experiment |
| **P6 - Creating** Generate new ideas, products, or viewpoints | Students **curate** work based on processes and experiences built from previous levels | ST revised their materials based on feedback and presented again to the group | Generates ideas through divergent thinking for students | *ChatGPT* can brainstorm to generate multiple different viewpoints | *DALL-E* can generate a series of unique and unlimited images to help students get inspiration for creative design |
| **Higher Dimensions** | Meta-cognitive process | | | | |

## 5. The Hierarchy of LBET

This paper establishes a new LLM-driven BET, which is shown in *Figure 2*, LBET continues to employ a pyramid structure with conditionally constrained linear sequences, with each level of classification building upon the previous one to deepen the cognitive process incrementally. This paper is oriented towards improving IL and uses the IR and problem-solving process to construct different levels. Some studies have discussed in depth the process of interaction between LLMs and students [62][63], which may have a positive impact on IL. Thus, a new dimension is added to more accurately describe the information interaction behaviors between higher education students and LLMs. These Phases are Perceiving, Searching, Reasoning, Interacting, Evaluating, Organizing, and Curating. This paper provides a more 'fine-grained' discussion of the process of using LLMs for IR and problem-solving: progressing from abstract concepts to specific student behaviors and integrating the use of LLMs as a tool to support student learning within the taxonomy to address **RQ1**, which will provide more details and discussion in the following sections.

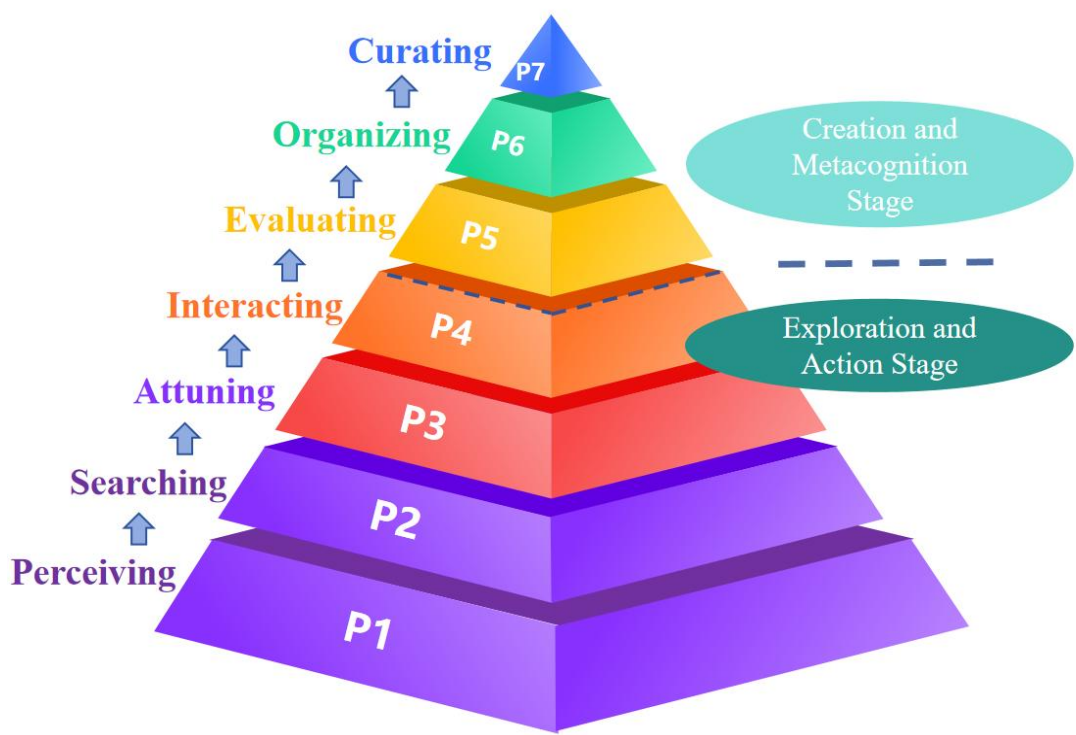

Figure 2. The structure of LBET. (There is an error in the image for 'Attuning', it will be corrected to 'Interacting' soon)

### 5.1 Perceiving

Perceiving is described as an objective, primary cognitive behavior that constitutes a component of exploratory learning and serves as a preliminary stage of action [64]. In LBET, perceiving specifically refers to the totality of students' initial objective recognition of both the model and the task when first engaging with LLMs. It is the foundational stage of LBET, laying the cornerstone for initiating the utilization journey of LLMs. Scenarios involving the use of LLMs are predominantly task-oriented [65]. In this stage, learners will draw upon their own experiences to judge and, through reasonable task decomposition, identify which aspects require the support of LLMs. They will explore how to use LLMs for preliminary IR and exploratory problem-solving. This perceiving can be non-logical, for example, when interdisciplinary interactions require the decomposition and independent consideration of involved disciplines without immediate demand for systematic organization. Additionally, since perceiving is related to the flow of life experiences and subjective attitudes toward an object [66], students will recognize that the functions of LLMs need to be dynamically integrated with their learning needs and personal experiences during this phase. Simultaneously, students will outline diverse solution paths for the task based on their varying disciplinary backgrounds and demographic characteristics. This personalized perceiving stage helps lay the foundation for HOTS, such as autonomous learning.

## 5.2 Searching

Searching is part of the broader concept of "action" serving as a pivotal element in any learning and cognitive process that involves IR systems within the LBET framework. Adequate preparation during the search phase is instrumental in optimizing subsequent stages of the process. Given that the ultimate task or objective is often intricate and abstract, the search process at this stage is inherently "exploratory"—without having a fully crystallized search objective from the outset. In this search process, students incrementally gather and examine pertinent information, continually refining and broadening their exploratory search strategies [67]. The effectiveness of these search strategies directly impacts the quality of cognition throughout the entire IR process and plays a critical role in facilitating precise retrieval during subsequent interactions [68]. Furthermore, as students engage with content across different modalities, accumulating sufficient depth of relevant background knowledge is essential. This foundational knowledge forms the raw material necessary for meaningful interaction and feedback in later stages. Failure to adequately explore or build upon these initial materials may lead to significant cognitive challenges, potentially necessitating a complete restart [69]. In more severe cases, such insufficient preparation can introduce biases during the evaluation phase, thereby undermining the overall integrity of the analysis.

## 5.3 Reasoning

Reasoning constitutes the second phase of LBET. It involves deducing conclusions through logical processes grounded in existing knowledge or data. This cognitive endeavor aims to analyze and integrate known information, data, or facts to generate new insights, conclusions, or determinations [70]. The reasoning trajectory typically includes analysis, judgment, induction, deduction, and abduction.

Conventional IR systems, such as Google, adhere to an "index-retrieve-rank" paradigm: users articulate their requirements via concise queries, prompting the system to select a subset of relevant results from the available document corpus, which are then ranked based on a specific algorithm (e.g., descending relevance [71]. Traditional systems generally do not aim to synthesize information from multiple documents into a unified response. In contrast, the generative retrieval paradigm of LLMs synthesizes all corpus information into a single model. Sequence-to-sequence models are trained to map queries directly to relevant document identifiers, often extending prompts into potential completions, delivering direct responses to information needs instead of directing users to primary sources [72]. However, this obfuscated retrieval process, which may surface more pertinent information while diminishing or negating interaction, could erode many foundational elements of search, including information verification, IL, and serendipity [73]. At the same time, it also inadvertently destroyed a mindset: searchers usually, but do not always, select the first result, even if a lower-down result is more relevant to their specific needs [74]. Consequently, when employing LLMs, students must not only engage in logical reasoning to make implicit retrieval and ranking processes explicit but also critically evaluate the synthesized responses to ensure they meet their individual information needs and foster higher-order cognitive abilities

It should be emphasized that the reasoning in LBET is ideally considered to be rational, but this is not always the case in reality. For instance, students' personal emotions and preferences, such as self-efficacy, play a positive role derived from Internet-based learning [75]. Students often consider both reasoning and emotion when addressing problems in practice. As such, attuning between the two is an important issue that lies beyond the scope of this paper but warrants further discussion.

### 5.4 Interacting

Interaction provides a way for students and LLMs to communicate and strengthen or reconstruct their knowledge based on specific IR strategies. Interaction is a dynamic iterative process: students interrogate the LLMs through such an IR strategy, the LLMs respond, and then through iterations, the students gradually build and expand on their formed knowledge system [76]. Most interactions with LLMs are in the form of text records [77][78][79]. One difference between the interactive level and reasoning is that the inquiries at this stage will form obvious interactive strategies and be executed, such as the berrypicking [80] to pursue and analyze the logical structure of a question [81]. It should be noted that students may not achieve a good result by interacting in a isolated way [69]. For instance, after students repeatedly iterate their inquiry into a question, the LLM will oscillate between the generated answers [82]. Also, another study showed that LLMs, although they may respond to their interlocutors' increasingly efficient speech, may not spontaneously become more efficient in their speech over time, as human students do [83]. To ensure a smooth and reliable process, students may attempt to employ diverse and multimodal approaches to ensure that LLMs accurately respond to their intentions or try prompt or fine-tune to make LLMs adapt to students' own interaction rhythm and style. This process, in turn, fosters their IL ability and is an integral part of the overall interaction.

In LBET, most levels progress sequentially, with one notable exception: the transition from interaction to evaluation can face obstacles. In use of LLMs, pre-trained databases and improper parameter adjustments may introduce irrelevant information [69]. Unlike traditional IR systems, where results are ranked by relevance, LLMs' opaque processes often lead to ambiguity and divergence in search outputs [84]. This implicit output mode enriches exploration but also increases the risk of deviation, requiring learners to possess higher cognitive skills [85]. Thus, students must frequently reassess and expand their perception and reasoning processes to manage risks and enhance learning effectiveness [86].

At this stage, students have gradually completed the transition from LOTS to HOTS and have acquired the prerequisites for achieving a variety of HOTS such as critical thinking ability: After gaining an understanding of a problem or task through searching, students gradually reconstruct and regenerate information to create new knowledge and concepts [87]. This stage is also when students receive the most information, and their cognitive processes are most active when they are searching for information.

## 5.5 Evaluating

Evaluating is a crucial component within LBET, serving to help students identify the strengths and weaknesses of their work. This includes reassessing large models' robustness [88], ethical considerations and biases [89][90], as well as credibility [91] and hallucinations [92] to prevent severe direct harm or unpredictable potential impacts stemming from inherent flaws in LLM outputs, while concurrently reinforcing cognitive and conceptual organization.

Evaluating methods for learners are diverse and can be broadly categorized into internal and external evaluations. Internal evaluation refers to calibration that occurs solely between the student and the LLMs. For example, students might repeatedly ask the same question to LLMs to test the robustness of the generated answers. Alternatively, more quantitative approaches, such as standardized metrics and evaluation tools like ROUGE scores [93], can be employed for horizontal comparison and evaluation of various models. External evaluation, on the other hand, involves obtaining corrections and supplementary feedback through channels beyond the learner and LLMs, including the use of various sources such as peer evaluations, teacher assessments, and other external channels [57]. These external supplements allow learners to examine LLM-generated results from multiple perspectives, thereby mitigating or avoiding additional ethical risks. The choice between these methods is based on objective conditions and practical considerations. In many cases, internal evaluation is a convenient, yet higher-risk approach and external evaluation typically involves inviting experts, researchers, or ordinary users to evaluate model-generated outputs in conjunction with real-world contexts, may require more effort but is generally safer and more cautious.

It is important to note that human evaluations can exhibit significant variance and instability, influenced by diverse cultural and individual factors, potentially resulting in varied human alignment [94]. In practical applications, a nuanced consideration and balance between these two evaluation approaches is required.

## 5.6 Organizing

Organizing represents a more advanced and intricate component within LBET. Organizing involves combining smaller tasks into a cohesive whole; students treat LLM-generated answers as documents, critically reading, examining, and synthesizing them within the LBET framework. Following evaluation, students develop a more comprehensive and detailed understanding of task-oriented nuances. Tasks often involve complexity and multidimensionality. For instance, completing a sophisticated IR task may require several sub-tasks across multiple rounds of searching and dialogue, involving numerous queries and interactions with various informational objects (e.g., documents and projects) [95]. The goal of organizing is to systematically arrange and synthesize these conceptual elements to enhance students' deeper understanding of overarching problems and theoretical frameworks. This represents the highest level of information utilization and processing. An example of organizing is when students put answers to different questions, they received on ChatGPT into a paragraph and describe them logically. Students often adhere to established methodologies such as systematic reviews or meta-analyses. LLMs can be leveraged to support these meta-analytic efforts [96].

## 5.7 Curating

Curating, in its broadest sense, refers to the selective display and presentation of materials in the digital age [97]. Curation is the process of refining previously developed ideas or models at a higher level, to create innovative solutions, such as those with enhanced visualization and other advanced features. Within our taxonomy, Curating represents the boundary of the physical completion of an information task. It necessitates that students step out of the narrow scope of dialogue with machines, and instead present their knowledge or outcomes in a larger context to seek help or support. For example, learners can seek professional comments from authoritative people or higher-level institutions or seek more diverse improvements from experts with interdisciplinary backgrounds.

The process of showcasing and sharing effectively reflects the necessity for students to expand from individual learning to communicating ideas widely via social and knowledge networks. The construction of LLMs is intrinsically driven by robust first-hand human historical data. However, as human-computer collaboration advances, it inevitably brings about issues such as data leakage and data contamination, leading to inherent limitations that may be difficult to resolve [98]. This could result in students facing unsolvable problems, making it essential for them to share their processes or outcomes across various channels—such as with peers, teachers, experts, and social networks—once they have completed a task or goal to facilitate further exploration and foster the growth of students' critical thinking skills.

## 6. The Stages of LBET

The corresponding stages in LBET are exploration and action, as well as creation and metacognition, as shown in Sections 6.1-6.2. While LBET is orientated towards by IL, it still contains the cognition stage of BET (LOTS and HOTS). These two stages correspond to students' IR ability and problem-solving ability in IL.

### 6.1 Exploration and Action

The Exploration and Action stage includes Perceiving, Searching, Reasoning, and Interacting. It emphasizes collecting and organizing resources related to specific topics or tasks, gradually shaping them into a coherent knowledge framework [99]. Perceiving and subsequent actions occur alternately, accompanied by exploration and information acquisition [100]. Through exploratory behavior, students actively engage and receive feedback through their interactions with LLMs, progressively advancing the learning process. Positive feedback motivates further rounds of actions involving repeated cycles of exploration and information gathering [101]. During this stage, students not only master LOTS—such as Remembering, Understanding, and Applying—but also continuously discover the "affordances" in integration with higher education students and LLMs, laying the foundation for HOTS.

### 6.2 Creation and Metacognition

Evaluating, Organizing, and Curating constitute the second stage, known as the Creation and Metacognition stage. Following the Exploration and Action stage, students progressively master HOTS, such as critical thinking and autonomous learning. They gradually develop the ability to handle complex problems and significantly improve their capacity to apply and utilize knowledge in LLMs. The phases of Evaluating, Organizing, and Curating represent a concrete manifestation of metacognition. Learners plan, monitor, and modify their cognitive activities to achieve a more efficient cognitive state. This state is often referred to as creative cognition.

Complex problems in practice typically require creativity to resolve, as they often lack a single correct solution that can be resolved through straightforward analytical methods [102]. Evaluating, Organizing, and Curating offers a structured approach to achieving creative cognition in ambiguous problem-solving. Learners refine their metacognitive processes to explore high-uncertainty problems from multiple perspectives, enabling them to critically develop unique insights and creatively propose solutions. Ultimately, students enhance their ability to autonomously solve complex problems through iterative cognitive cycles and creative thinking [103].

## 7. Case Study of LBET

Observing students' use of LLMs was an exploratory and forward-looking project. We conducted a small observational experiment and observed the effectiveness of LBET in answering **RQ2** in assessing and guiding students to use LLMs and their IL levels in terms of applicability and feasibility. The case investigation comprised three stages: pre-test, observation experiment, and post-test, as shown in *Figure 3.*

The pre-test consisted of a brief interview conducted before the formal observation experiment to understand the participants' experience with LLMs and their data analysis background. Based on the interview results, a preliminary assessment was made regarding which stage of the LBET framework each participant might correspond to. Subsequently, the same task and dataset were provided to all participants, who were allowed to use designated LLMs to complete the task. More details on the experimental process are in Appendix A.

After the observation experiment, the post-test was conducted. It included a re-observation experiment and re-evaluation, with the LBET framework provided to students who paused their participation to support task continuation. This was followed by a semi-structured interview with fixed questions and by asking students to describe their ideas and steps in using LLMs to verify the accuracy of the observations. More details and results are shown in *Appendix B.*

It should be emphasized that these examples serve as an introduction to the framework's feasibility. Although various issues and findings in the students' behaviors and strategies within LBET were revealed, a comprehensive explanation of how best to address these issues is necessary. Unfortunately, analyzing the causes of these differences exceeds the scope of this paper. In considering the 'practicality' of addressing **RQ2**, our research case has reached saturation in terms of conventional qualitative research [104].

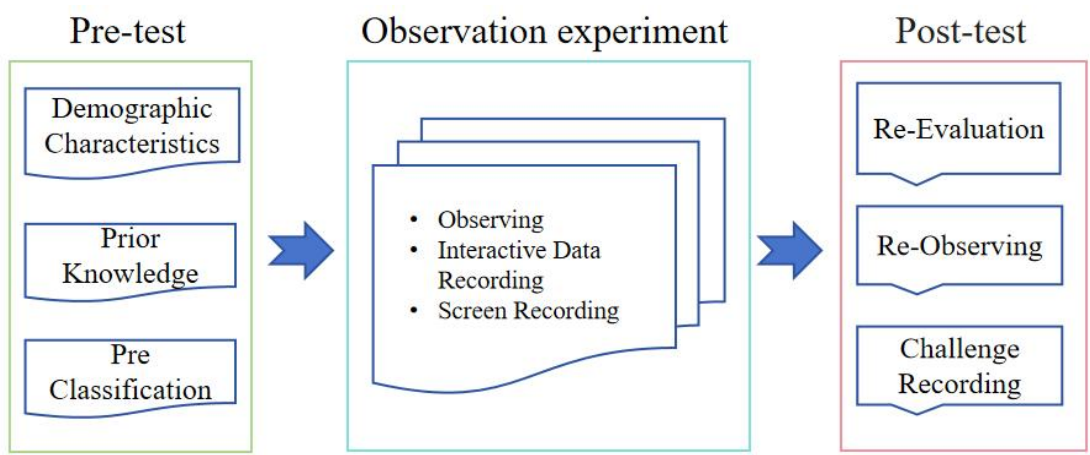

Figure 3. Flow chart of feasibility experiment.

## 7.1 Applicability of the LBET Framework

The preliminary survey information for the five participants in the pre-test is presented in *Table* 2, with all names being pseudonyms. The participants displayed diverse prior conditions and exhibited a range of progressive attitudes toward the use of LLMs. This diversity effectively illustrated the framework's adaptability and potential for broad applicability across varying conditions. Also visualized in *Figure 4,* we conducted an observation experiment on the participants and classified their LLM application levels through LBET. Afterwards, all the data were submitted to five researchers who were PhD candidates with relevant AI and education backgrounds for classification using LBET. The hierarchy was determined by the principle of the majority overriding the minority. The results showed that the LBET framework could clearly describe and distinguish the use of LLMs by all students, supporting the framework's applicability to a certain extent.

The LBET clearly reveals that:
1. *Ann* experienced a halt in progress after the searching behavior and subsequently withdrew from the test.
2. *Bob* became disoriented during continuous inquiries as GPT's output increasingly diverged into irrelevant and erroneous topics, eventually leading to the student's withdrawal from the test.
3. *John* completed the test after completing the evaluation of a statistical task based on the uncertainty of the final integrated task.
4. *Frank* submitted all the original documents, which included redundant and irrelevant content.
5. *Lily* successfully completed the test.

In the post-test, we showed our framework description of Section 5 to the three students who dropped out and asked them to continue to perform the task. The task situation is as follows:
1. *Ann* asked GPT multiple times based on the same question, then chose an answer that seemed more reasonable, and then gave up again.
2. *Bob* collected different answers, showed the same prompt word project at the same time and asked GPT to point out the differences, and repeatedly asked if there were any errors. In the process of testing robustness by repeatedly asking GPT the same question with different prompts, *Bob* discovered that during the data exploration process, GPT had made various modifications to the predictive algorithms. After further inquiry, he ultimately found that the machine learning algorithms provided were not suitable for the current subtask involving descriptive data statistics.
3. *John* reopened a new round of inquiries, put the results of multiple subtasks together in the document and actively selected them.
4. *Frank* reorganized the content and introduced detailed PowerPoint for curating.

In the post-test, according to the experts' categorical ratings again based on the LBET, it was observed that students in all previous conditions had improved their problem-solving abilities using LLM to varying degrees, as shown by the pink marks in *Figure 4*. Also, most students agreed that the framework accurately categorized their LLM usage and said that it enhanced their understanding of how to use LLM for learning, as shown in *Appendix B*. This reflects that the framework has a certain positive effect on standardizing and guiding students to use LLMs to solve complex problems.

**Table 2.  Demographic Information**

| Case | Gender | Age | Departments | Knowledge of LLMs | Opinion | Reasons for Usage | Prior Knowledge* |
|------|--------|-----|-------------|-------------------|---------|-------------------|------------------|
| Ann | Female | 19 | Science and Technology | Heard of LLMs but no usage experience | Feels LLMs are versatile but unhelpful for useful information; prefers writing assignments independently | Difficulty finding an interface; distrust; believes it's inefficient | No (0) |
| John | Male | 18 | Business Administration | Heard of LLMs, has usage experience, asks when unsure | Distrustful; sometimes fails to deliver good results; considers it limited; uses multiple AI tools | Helps to explore ideas; better search engine compared to traditional ones; timesaving | A bit (1) |
| Frank | Male | 18 | Science and Technology | Heard of LLMs and have a few usages experience | Distrustful; feels it is not useful | Device does not support it; if supported, would use it mainly for retrieving information | Know Programming (3) |
| Lily | Female | 19 | Science and Technology | Heard of LLMs, proficient in using LLMs | Neutral, needs more observation | Use LLMs for retrieval, translation, and organization | Understand how it works, Able to use skillfully (5) |
| Bob | Male | 19 | Business Administration | Heard of LLMs, has usage experience, uses critically | Distrustful; believes it doesn't solve some problems; prefers alternative solutions | Helps to explore ideas; better search engine compared to traditional ones; timesaving | No (0) |

*Prior Knowledge is based on a three-point Likert scale

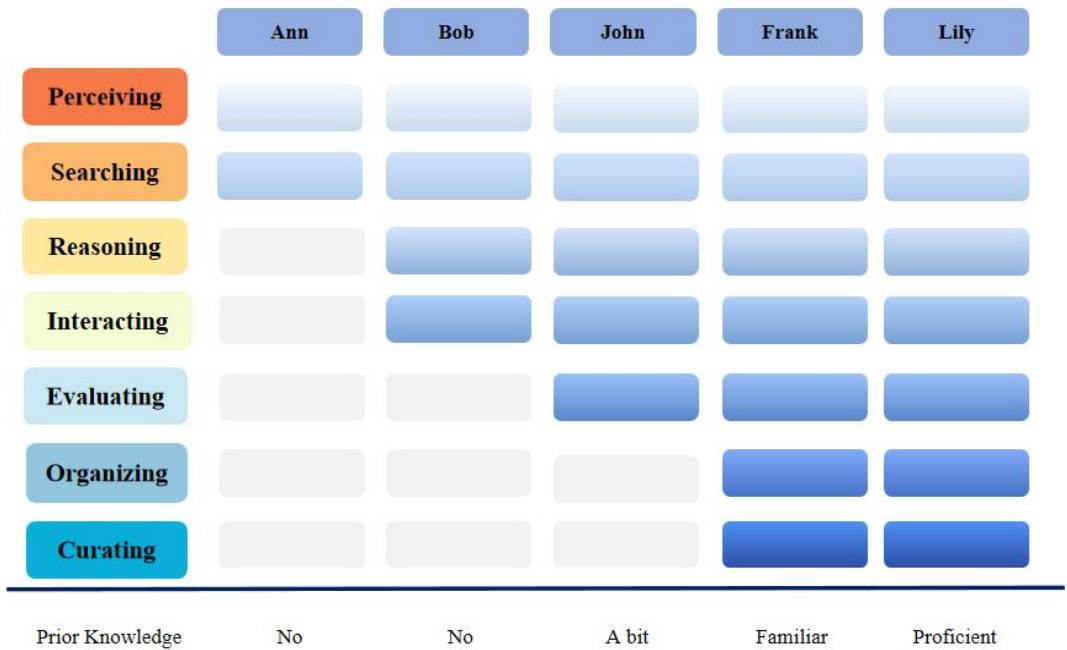

**Figure 4. Levels of different students in LBET. The taxonomic ratings for the pre-test are indicated by blue markings, and the taxonomy ratings for the post-test are indicated by red markings (Red marks are not displayed correctly and will be corrected soon).**

### 7.2 Feasibility of the LBET framework

A case study was taken, and *Lily* was selected as the test subject to conduct a preliminary exploration of the feasibility of the LBET framework, following purposive sampling strategies [105]. *Lily* was chosen because after fully completing the test, the research found that her results can show the complete process of IR and problem-solving, ensuring the depth and quality of the data. *Lily* understands the principles behind LLMs and has a certain level of experience using them. *Lily* maintains a neutral attitude towards LLMs. Her cognitive process is illustrated in ***Figure 5*** and can be found in more detail in Appendix C. *Lily* successfully completed the task by leveraging her knowledge and interacting with GPT, exemplifying how the LBET framework can effectively capture her use of LLMs and offer a viable case study.

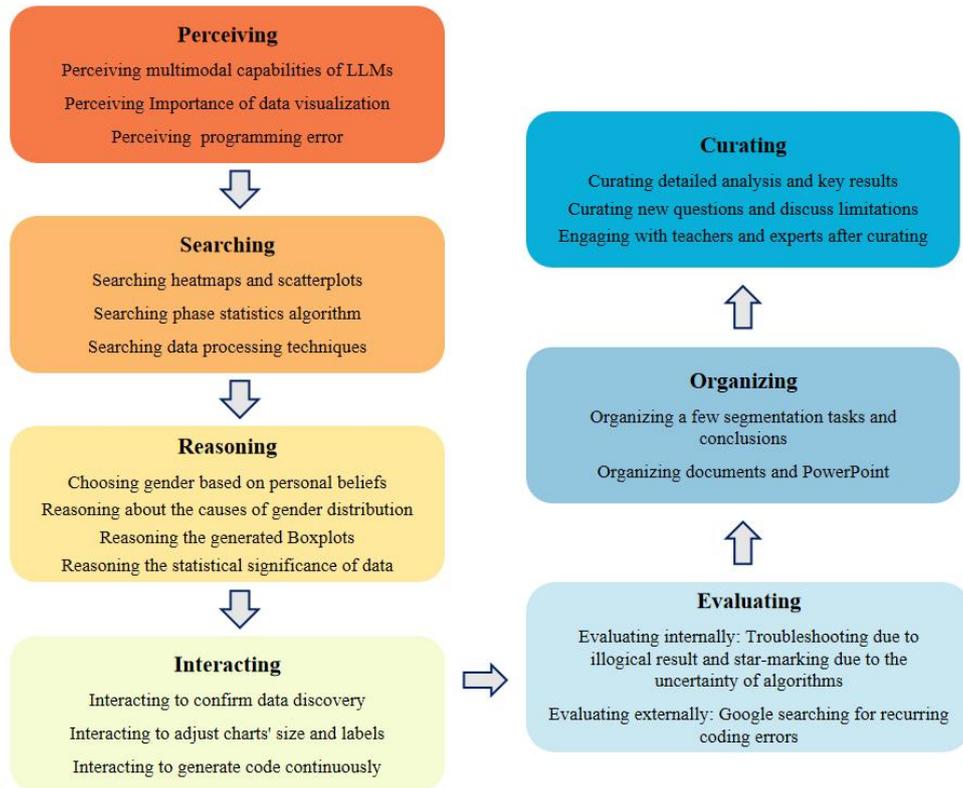

**Figure 5.** *Lily*'s flow chart of using LLMs to solve problems. This flow chart contains the key behaviors of *Lily*'s thinking based on the LBET model.

### 7.3 Limitations

The LBET framework offers a parallel ascending structure, where each level presents students with different challenges. It can be foreseen in the feasibility examples that each stage may contain many valuable insights. Analyzing these behaviors and cognition-related processes requires more in-depth observational experiments to provide further insight and discussion. Undoubtedly, by utilizing the LBET framework, we have identified a structured pathway that guides and facilitates deeper learning in human-computer interaction with LLMs, even in the absence of teacher intervention.

However, it is noteworthy that preliminary observations from the experimental study indicate that students face challenges in actively discovering and exploring more advanced stages. This difficulty may arise partly from the fine granularity of observational study. For example, in the post-test, *John*, *Frank*, and *Lily* have reached the highest level in taxonomy, but t this does not imply that their final curated outputs would receive equivalent scores: *John* submitted a Word document with key points in text form, while *Frank* submitted a PowerPoint presentation with images and detailed content. *Lily* provided a PowerPoint file that only included an abstract and outline, but she also included video links and multiple academic references. While they exhibit some differences in visualization and presentation, further subdivision of evaluation at the horizontal level remains elusive due to the absence of clearly defined criteria for assessing quality at specific stages within the IL framework.

In future work, more empirical analysis will be conducted to demonstrate the framework's applicability in different situations and to further validate and improve it. Additionally, potential issues and assumptions will be examined more precisely. The research will also establish standards for further exploring the completion of specific stages within IL skills.

## 8. Conclusion

The practice of LLMs is reshaping how students approach complex tasks. This paper introduces an LLM-driven Bloom Educational Taxonomy, LBET: a novel framework that encapsulates the fusion of LLMs and students' IL. By examining students' IR behaviors and cognitive processes through the seven dimensions of LBET, the framework provides educators with a comprehensive tool for regulating and guiding students in the effective use of LLMs, and also shows a positive path to qualitatively analyze students' "search quotas" in the LLMs era. Simultaneously, we conducted a framework feasibility presentation using a pre-test, observational experiment, and post-test case analysis model. Through our case study, we have found that LBET is well-suited to students with diverse prior conditions. Additionally, our analysis has revealed that implementing this framework in the post-test stage positively impacted problem-solving abilities across various stages of students. This provides support for the framework's potential to foster students' IL. Our framework not only contributes to a deeper understanding of how to effectively utilize LLMs but also offers a structured approach to fostering IR and problem-solving abilities in the modern information age. Looking forward, the LBET can be a tool with significant contribution for observing IL with LLMs and its impact on cognitive development.

## Appendix A

**Experiment description**

We conducted an observational experiment of "LLM-based IR and task completion" with five second-year students from a public university in Zhuhai, China. All five students were enrolled in a data science elective, so this experiment will be conducted in accordance with the data science curriculum and assignment outline set by the university. The task was described as follows: "You will be provided with an Excel dataset containing various student characteristics and academic performance information. Please analyze the dataset to uncover insights regarding the relationship between student characteristics and performance and organize your findings for presentation."

The dataset used was a publicly available Kaggle student dataset, and ChatGPT-4o was designated as the LLM for the experiment (we used **GPT** to refer to this LLM hereafter). Each person's GPT account is independent and repartitioned separately to avoid the influence of prior memories. The participants were isolated in separate and enclosed rooms. Before the experiment, researchers conducted a brief interview with the students to collect basic information including demographics and attitudes toward LLMs. In addition, a brief introduction to data science was given to participants. After being briefly introduced to data science and LLMs, the experimenters conducted one hour of behavioral observation. Afterwards, a new round of interviews was conducted with the students.

# Appendix B

In the post-test phase, for students who gave up or paused the task, we provided the LBET framework to help them continue to complete the task, and the students were in place to ensure that they could move forward to completion. Judges continued to categorize their IL level using LLMs based on typical behaviors. The results of the fixed questions in the semi-structured interview are shown in *Table B*.

**Table B. Post-test Results**

| Case | Why did you give up on testing? | Do you feel the framework accurately categorizes your stages? * | Does the framework increase your confidence in using LLMs? * | Will you use LLMs more in the future? * |
|---|---|---|---|---|
| Ann | I don't know how to use it | Neutral | Neutral | Yes |
| John | I don't know how to organize my results | Yes | Yes | Yes |
| Frank | / | Yes | Yes | Yes |
| Lily | / | Yes | Yes | Yes |
| Bob | The GPT output is confusing me, but I don't know what to do | Yes | Yes | Yes |

**\*Questions are based on a three-point Likert scale**

## Appendix C

Appendix C shows a text description of *Lily's* information behavior as observed by the researchers. This text description is reorganized based on all original audio and web page records.

**1. Perceiving**

*Lily* uploaded both the problem and dataset to GPT for analysis after receiving the task. GPT, using Python code functionality, provided constructive suggestions, including exploring the relationships between extracurricular activities, study habits, and academic performance. Realizing that data visualization would be an effective approach, *Lily* asked GPT whether it could present the data using scatter plots and received a recommendation to further explore the data with scatter plots, heat maps, and bar charts. Aware of the potential need for coding, *Lily* instinctively opened the locally installed Python software (PyCharm) and chose to directly copy the code for local execution, thus avoiding potential errors from missing local libraries that GPT might have encountered.

**2 Searching**

As *Lily* was not familiar with the specifics of heat maps and scatter plots, they relied on GPT to search for relevant knowledge. GPT provided definitions, labels, and advantages of different charts, including correlation coefficients and regression analysis. *Lily* continued to inquire about these related concepts while adding more practical questions, such as how to handle the dataset using these concepts and how to visualize the data through coding.

**3 Reasoning**

Gradually, *Lily* began to record, organize, and categorize the information they had retrieved. Through this process, she confirmed the specific insights they intended to present: an exploration of the gender distribution across different subjects. As Lily researched the gender distribution of various subjects, she realized that her initial motivation stemmed from her personal belief in the importance of gender equality in education. By this stage, *Lily* shifted from exploratory searching to more precise inquiries, asking GPT to generate box plots for various categories and to use p-values to test whether there were differences in performance across subjects between male and female students.

**4. Interacting**

During the Interacting phase, *Lily* continuously refined task performance by engaging in prompt engineering with GPT, using commands to modify code and improve the visuals, including adjusting the size, labels, and values of the graphs to enhance their aesthetic appeal. Additionally, when encountering errors in the locally executed code, *Lily* uploaded the results to GPT for troubleshooting, prompting GPT to offer alternative solutions, such as self-written code or the use of other Python libraries to resolve the issue.

**5. Evaluating**

In the evaluation phase, *Lily* adopted an internal evaluation approach to verify and review the results. When predicting academic performance based on future career aspirations, the prediction performance yielded only a 30% recall rate. This prompted *Lily* to further query GPT, asking it to trace back its previous decisions step by step to identify potential sources of the issue. It was eventually discovered that GPT had used a randomly generated sample of data during a previous demonstration, rather than the actual dataset. In addition, *Lily* was wary of the prediction method provided by the machine learning algorithm provided by GPT. Because of the lack of prior knowledge, they finally started the method and used Google to query it and found that it was contrary to their original intentions. The plan was later abandoned.

### 6. Organizing

At this stage, *Lily* began to initiate new sub-tasks and integrate multiple subtasks into a complete project file. *Lily* continued to use GPT to generate various outputs, including documents, code, tables, and images, and asked GPT to create a PowerPoint presentation while making fine adjustments to text size, image layout, and other elements. *Lily* completed the organizational structure of the presentation, including sections on the problem, findings, discussion, and results.

### 7. Curating

*Lily* selectively presented the results in a PowerPoint oral report to the researchers, providing a detailed analysis and selective presentation of the content. In curating the exhibition, she used additional sources including YouTube videos and academic literature to make the visual expression stronger. Based on the results of the data analysis, *Lily* proposed new problems and initiated reasonable discussions. *Lily* gave a brief presentation on the relationship between student grades and gender but focused more on the influence of future career aspirations on academic achievement. *Lily* hypothesized that career choice might have a greater impact on academic performance than other factors, followed by a discussion of potential reasons and a summary of findings.

## Declarations
*Ethics approval and consent to participate*
   Not applicable.
*Consent for publication*
   Not applicable.
*Availability of data and materials*
   The author confirms that all data generated or analysed during this study are included in this published article. Furthermore, primary and secondary sources and data supporting the findings of this study were all publicly available at the time of submission.
*Competing interests*
   The authors declare no competing interests.
*Funding*
   This research was funded by Macao Science and Technology Development Fund (funding ID: 0088/2023/ITP2) and Macao Polytechnic University research grant (project code: RP/FCA-10/2022).
*Authors' contributions*

Y.L. conceptualized and designed the study as well as prepared the manuscript. Y.L. and T.L. conducted the experiments and data collection. P.P. supervised the overall execution of the study and acquired funding for the research. D.M., Z.C., G.B., and S.C. reviewed and edited the manuscript.

## References

[1] Liu, H., Dai, Z., So, D., & Le, Q. V. (2021). Pay attention to mlps. *Advances in neural information processing systems, 34*, 9204-9215.

[2] Wang, X., Pang, H., Wallace, M. P., Wang, Q., & Chen, W. (2024). Learners' perceived AI presences in AI-supported language learning: A study of AI as a humanized agent from community of inquiry. *Computer Assisted Language Learning*, *37*(4), 814-840.

[3] Urban, M., Děchtěrenko, F., Lukavský, J., Hrabalová, V., Svacha, F., Brom, C., & Urban, K. (2024). ChatGPT improves creative problem-solving performance in university students: An experimental study. *Computers & Education*, *215*, 105031.

[4] Murgatroyd, S. (2024). The Future of Higher Education in an Age of Artificial Intelligence. Ethics International Press.

[5] Dziuban, C., McMartin, F., Morgan, G., Morrill, J., Moskal, P., & Wolf, A. (2013). Examining student information seeking behaviors in higher education. *Journal of Information Fluency, 2*(1), 36-54.

[6] Snavely, L., & Cooper, N. (1997). The information literacy debate. *The Journal of Academic Librarianship*, 23(1), 9-14.

[7] Connaway, L. S., Julien, H., Seadle, M., & Kasprak, A. (2017). Digital literacy in the era of fake news: Key roles for information professionals. *Proceedings of the Association for Information Science and Technology*, *54*(1), 554-555.

[8] American Library Association. (1989). American library association presidential committee on information literacy. *http://www. ala. org/ala/acrl/acrlpubs/whitepapers/presidential. htm.*

[9] Norman, G. R. (1988). Problem‐solving skills, solving problems and problem‐based learning. *Medical education, 22*(4), 279-286.

[10] Fosnacht, K. (2017). Information literacy's influence on undergraduates' learning and development: Results from a large multi-institutional study.

[11] Al-Zou'bi, R. (2021). The impact of media and information literacy on acquiring the critical thinking skill by the educational faculty's students. *Thinking Skills and Creativity*, 39, 100782.

[12] Kasneci, E., Seßler, K., Küchemann, S., Bannert, M., Dementieva, D., Fischer, F., ... & Kasneci, G. (2023). ChatGPT for good? On opportunities and challenges of large language models for education. *Learning and individual differences*, 103, 102274.

[13] Filippi, S., & Motyl, B. (2024). Large Language Models (LLMs) in Engineering Education: A Systematic Review and Suggestions for Practical Adoption. *Information, 15*(6), 345.

[14] Askarbekuly, N., & Aničić, N. (2024). LLM examiner: automating assessment in informal self-directed e-learning using ChatGPT. *Knowledge and Information Systems*, 1-18.

[15] Henkel, O., Hills, L., Boxer, A., Roberts, B., & Levonian, Z. (2024, July). Can Large Language Models Make the Grade? An Empirical Study Evaluating LLMs Ability To Mark Short Answer Questions in K-12 Education. In P*roceedings of the Eleventh ACM Conference on Learning@ Scale* (pp. 300-304).

[16] Gao, Y., Wang, Q., & Wang, X. (2024). Exploring EFL university teachers' beliefs in integrating

ChatGPT and other large language models in language education: a study in China. *Asia Pacific Journal of Education*, 44(1), 29-44.

[17] Gao, Y., Wang, Q., & Wang, X. (2024). Exploring EFL university teachers' beliefs in integrating ChatGPT and other large language models in language education: a study in China. *Asia Pacific Journal of Education*, 44(1), 29-44.

[18] Jomuad, P. D., Antiquina, L. M. M., Cericos, E. U., Bacus, J. A., Vallejo, J. H., Dionio, B. B., ... & Clarin, A. S. (2021). Teachers' workload in relation to burnout and work performance. *International journal of educational policy research and review*.

[19] Bouchard, G. J. (2011). In Full Bloom: Helping Students Grow Using the Taxonomy of Educational Objectives. *The Journal of Physician Assistant Education*, 22(4), 44-46.

[20] Chandio, M. T., Pandhiani, S. M., & Iqbal, R. (2016). Bloom's Taxonomy: Improving Assessment and Teaching-Learning Process. *Journal of education and educational development*, *3*(2), 203-221.

[21] Metfessel, N. S., Michael, W. B., & Kirsner, D. A. (2012). Instrumentation of Bloom's and Krathwohl's taxonomies for the writing of educational objectives. In *Educational Objectives and the Teaching of Educational Psychology* (pp. 207-213). Routledge.

[22] Rahanu, H., Georgiadou, E., Khan, N., Colson, R., Hill, V., & Edwards, J. A. (2016). The development of student learning and information literacy: A case study. *Educ. Inf.*, *32*(3), 211-224.

[23] Meyer, J., Jansen, T., Schiller, R., Liebenow, L. W., Steinbach, M., Horbach, A., & Fleckenstein, J. (2024). Using LLMs to bring evidence-based feedback into the classroom: AI-generated feedback increases secondary students' text revision, motivation, and positive emotions. *Computers and Education: Artificial Intelligence*, 6, 100199.

[24] Dempere, J., Modugu, K., Hesham, A., & Ramasamy, L. K. (2023, September). The impact of ChatGPT on higher education. In *Frontiers in Education* (Vol. 8, p. 1206936). Frontiers Media SA.

[25] Fuchs, K. (2023, May). Exploring the opportunities and challenges of NLP models in higher education: is Chat GPT a blessing or a curse?. In *Frontiers in Education* (Vol. 8, p. 1166682). Frontiers Media SA.

[26] Besharat-Mann, R. (2024). Can I trust this information? Using adolescent narratives to uncover online information seeking processes. *Journal of Media Literacy Education*, 16(1), 1-18.

[27] Hoeber, O., & Storie, D. (2022, June). Information seeking within academic digital libraries: a survey of graduate student search strategies. In P*roceedings of the 22nd ACM/IEEE Joint Conference on Digital Libraries* (pp. 1-5).

[28] Kazemitabaar, M., Hou, X., Henley, A., Ericson, B. J., Weintrop, D., & Grossman, T. (2023, November). How novices use LLM-based code generators to solve CS1 coding tasks in a self-paced learning environment. In *Proceedings of the 23rd Koli Calling International Conference on Computing Education Research* (pp. 1-12).

[29] Stadler, M., Bannert, M., & Sailer, M. (2024). Cognitive ease at a cost: LLMs reduce mental effort but compromise depth in student scientific inquiry. Computers in Human Behavior, 160, 108386.

[30] Dempere, J., Modugu, K., Hesham, A., & Ramasamy, L. K. (2023, September). The impact of ChatGPT on higher education. In *Frontiers in Education* (Vol. 8, p. 1206936). Frontiers Media SA.

[31] Sharma, N., Liao, Q. V., & Xiao, Z. (2024, May). Generative Echo Chamber? Effect of

LLM-Powered Search Systems on Diverse Information Seeking. In *Proceedings of the CHI Conference on Human Factors in Computing Systems* (pp. 1-17).

[32] Hebenstreit, K., Praas, R., & Samwald, M. (2023). A collection of principles for guiding and evaluating large language models. *arXiv preprint arXiv:2312.10059*.

[33] Floridi, L. (2023). The ethics of artificial intelligence: Principles, challenges, and opportunities.

[34] Barman, K. G., Wood, N., & Pawlowski, P. (2024). Beyond transparency and explainability: on the need for adequate and contextualized user guidelines for LLM use. *Ethics and Information Technology, 26(3), 47.*

[35] Carroll, A. J., & Borycz, J. (2024). Integrating large language models and generative artificial intelligence tools into information literacy instruction. *The Journal of Academic Librarianship, 50(4), 102899.*

[36] Kathwohl, D. R. (2002). A Revision of Bloom's Taxonomy: An Overview. *Theory Into Practice, 41(4), 212-218*

[37] Wang, L., Chen, X., Deng, X., Wen, H., You, M., Liu, W., ... & Li, J. (2024). Prompt engineering in consistency and reliability with the evidence-based guideline for LLMs. *npj Digital Medicine, 7(1), 41.*

[38] Xiao, C., Xu, S. X., Zhang, K., Wang, Y., & Xia, L. (2023, July). Evaluating reading comprehension exercises generated by LLMs: A showcase of ChatGPT in education applications. In *Proceedings of the 18th Workshop on Innovative Use of NLP for Building Educational Applications (BEA 2023)* (pp. 610-625).

[39] McKay, D., Makri, S., Gutierrez-Lopez, M., MacFarlane, A., Missaoui, S., Porlezza, C., & Cooper, G. (2020, March). We are the change that we seek: information interactions during a change of viewpoint. *In Proceedings of the 2020 Conference on Human Information Interaction and Retrieval* (pp. 173-182).

[40] Shah, C., & Bender, E. M. (2022, March). Situating search. In *Proceedings of the 2022 Conference on Human Information Interaction and Retrieval* (pp. 221-232).

[41] Gürcan, Ö. (2024). LLM-Augmented Agent-Based Modelling for Social Simulations: Challenges and Opportunities. *HHAI 2024: Hybrid Human AI Systems for the Social Good,* 134-144.

[42] Zhang, Z., Zhang-Li, D., Yu, J., Gong, L., Zhou, J., Liu, Z., ... & Li, J. (2024). Simulating classroom education with llm-empowered agents. *arXiv preprint arXiv:2406.19226.*

[43] Bakas, N. P., Papadaki, M., Vagianou, E., Christou, I., & Chatzichristofis, S. A. (2023, December). Integrating LLMs in Higher Education, Through Interactive Problem Solving and Tutoring: Algorithmic Approach and Use Cases. In *European, Mediterranean, and Middle Eastern Conference on Information Systems* (pp. 291-307). Cham: Springer Nature Switzerland.

[44] Zhang, Y., Wang, X., Chen, H., Fan, J., Wen, W., Xue, H., ... & Zhu, W. (2024, August). Large Language Model with Curriculum Reasoning for Visual Concept Recognition. In *Proceedings of the 30th ACM SIGKDD Conference on Knowledge Discovery and Data Mining* (pp. 6269-6280).

[45] Westerlund, M., & Shcherbakov, A. (2024, March). LLM Integration in Workbook Design for Teaching Coding Subjects. In *International Conference on Science and Technology Education* (pp. 77-85). Cham: Springer Nature Switzerland.

[46] Cai, L., Jin, M., & Zhou, Y. (2023). The Essence of Education Digital Transformation: From Technology Integration to Human-machine Fusion. *Journal of East China Normal University (Educational Science*s), 41(3), 36.

[47] Bloom, B. S., Engelhart, M. D., Furst, E. J., Hill, W. H., & Krathwohl, D. R. (1964). Taxonomy of

educational objectives (Vol. 2). New York: Longmans, Green.

[48] Wilson, L. O. (2016). Anderson and Krathwohl – Blooms taxonomy revised. *Understanding the new version of Bloom's taxonomy*.

[49] Bach, T. A., Khan, A., Hallock, H., Beltrão, G., & Sousa, S. (2024). A systematic literature review of user trust in AI-enabled systems: An HCI perspective. *International Journal of Human–Computer Interaction*, 40(5), 1251-1266.

[50] Furst, E. J. (1981). Blooms taxonomy of educational objectives for the cognitive domain: Philosophical and educational issues. *Review of educational research*, 51(4), 441-453.

[51] Doherty, V. W., & Hathaway, W. E. (1972). Goals and Objectives in Planning and Evaluation: A Second Generation. *NCME Measurement in Education*. Vol. 4, No. 1, Fall 1972.

[52] Forehand, M. (2010). Bloom's taxonomy. *Emerging perspectives on learning, teaching, and technology*, 41(4), 47-56.

[53] Darwazeh, A. N. (2016). A rationale for revising Blooms [revised] taxonomy. thannual, 197.

[54] Dabić, T. (2016). The Bloom's taxonomy revisited in the context of online tools. In *Sinteza 2016-International Scientific Conference on ICT and E-Business Related Research* (pp. 315-320). Singidunum University.

[55] Churches, A. (2010). Bloom's digital taxonomy.

[56] Kilipiris, F., Avdimiotis, S., Christou, E., Tragouda, A., & Konstantinidis, I. (2024). Blooms Taxonomy Student Persona Responses to Blended Learning Methods Employing the Metaverse and Flipped Classroom Tools. *Education Sciences*, 14(4), 418.

[57] Demir, M. (2024). A taxonomy of social media for learning. Computers & Education, 105091.

[58] Norman, E., Pfuhl, G., Sæle, R. G., Svartdal, F., Låg, T., & Dahl, T. I. (2019). Metacognition in psychology. *Review of General Psychology*, 23(4), 403-424.

[59] Faraon, M., Granlund, V., & Rönkkö, K. (2023, September). Artificial Intelligence Practices in Higher Education Using Bloom's Digital Taxonomy. In *2023 5th International Workshop on Artificial Intelligence and Education (WAIE)* (pp. 53-59). IEEE.

[60] Achiam, J., Adler, S., Agarwal, S., Ahmad, L., Akkaya, I., Aleman, F. L., ... & McGrew, B. (2023). Gpt-4 technical report. *arXiv preprint arXiv:2303.08774.*

[61] Betker, J., Goh, G., Jing, L., Brooks, T., Wang, J., Li, L., ... & Ramesh, A. (2023). Improving image generation with better captions. *Computer Science. https://cdn. openai. com/papers/dall-e-3. pdf*, 2(3), 8.

[62] Salminen, J., Jung, S. G., Medina, J., Aldous, K., Azem, J., Akhtar, W., & Jansen, B. J. (2024, July). Using Cipherbot: An Exploratory Analysis of Student Interaction with an LLM-Based Educational Chatbot. In *Proceedings of the Eleventh ACM Conference on Learning@ Scale* (pp. 279-283).

[63] Bernabei, M., Colabianchi, S., Falegnami, A., & Costantino, F. (2023). Students' use of large language models in engineering education: A case study on technology acceptance, perceptions, efficacy, and detection chances. *Computers and Education: Artificial Intelligence*, 5, 100172.

[64] Gibson, E. J. (1988). Exploratory behavior in the development of perceiving, acting, and the acquiring of knowledge. *Annual review of psychology*, 39(1), 1-42.

[65] Chen, C., Yao, B., Ye, Y., Wang, D., & Li, T. J. J. (2024, May). Evaluating the LLM Agents for Simulating Humanoid Behavior. *The First Workshop on Human-Centered Evaluation and Auditing of Language Models* (CHI Workshop HEAL).

[66] Sun, M., Wang, M., & Wegerif, R. (2019). Using computer‐based cognitive mapping to improve

students' divergent thinking for creativity development. *British Journal of Educational Tchnology*, 50(5), 2217-2233.

[67] van Haastrecht, M., Sarhan, I., Yigit Ozkan, B., Brinkhuis, M., & Spruit, M. (2021). SYMBALS: A systematic review methodology blending active learning and snowballing. *Frontiers in research metrics and analytics*, 6, 685591.

[68] Pang, P. C. I., Verspoor, K., Chang, S., & Pearce, J. (2015). Conceptualising health information seeking behaviours and exploratory search: result of a qualitative study. *Health and Technology*, 5, 45-55.

[69] Luo, Y., Pang, P. C. I., & Chang, S. (2025). Enhancing exploratory learning through exploratory search with the emergence of large language models. *Proceedings of the 58th Hawaii International Conference on System Sciences* (HICSS 2025), Big Island, HI, USA, January 7-10, 2025(Accepted). *arXiv preprint arXiv:2408.08894.*

[70] Yu, F., Zhang, H., Tiwari, P., & Wang, B. (2023). Natural language reasoning, a survey. *ACM Computing Surveys.*

[71] Najork, M. (2023, July). Generative information retrieval. In Proceedings of the *46th International ACM SIGIR Conference on Research and Development in Information Retrieval* (pp. 1-1).

[72] Tang, Y., Zhang, R., Guo, J., & de Rijke, M. (2023, November). Recent advances in generative information retrieval. In *Proceedings of the Annual International ACM SIGIR Conference on Research and Development in Information Retrieval in the Asia Pacific Region* (pp. 294-297).

[73] Shah, C., & Bender, E. M. (2022, March). Situating search. In *Proceedings of the 2022 Conference on Human Information Interaction and Retrieval* (pp. 221-232).

[74] Joachims, T., Granka, L., Pan, B., Hembrooke, H., & Gay, G. (2017, August). Accurately interpreting clickthrough data as implicit feedback. In *Acm Sigir Forum* (Vol. 51, No. 1, pp. 4-11). New York, NY, USA: Acm.

[75] Tsai, C. C., Chuang, S. C., Liang, J. C., & Tsai, M. J. (2011). Self-efficacy in Internet-based learning environments: A literature review. J*ournal of Educational Technology & Society*, 14(4), 222-240.

[76] Rachha, A. K. (2024). Incorporating LLM-based Interactive Learning Environments in CS Education: Learning Data Structures and Algorithms using the Gurukul platform (Doctoral dissertation, Virginia Tech).

[77] Mandvikar, S. (2023). Augmenting intelligent document processing (idp) workflows with contemporary large language models (llms). *International Journal of Computer Trends and Technology*, 71(10), 80-91.

[78] Joko, H., Chatterjee, S., Ramsay, A., De Vries, A. P., Dalton, J., & Hasibi, F. (2024, July). Doing Personal LAPS: LLM-Augmented Dialogue Construction for Personalized Multi-Session Conversational Search. In *Proceedings of the 47th International ACM SIGIR Conference on Research and Development in Information Retrieval* (pp. 796-806).

[79] Dehbozorgi, N., Kunuku, M. T., & Pouriyeh, S. (2024, July). Personalized Pedagogy Through a LLM-Based Recommender System. In *International Conference on Artificial Intelligence in Education* (pp. 63-70). Cham: Springer Nature Switzerland.

[80] Jie, L., Sisi, G., & Xiaojuan, Z. (2023). Automatic Recognition of Exploratory and Lookup Intents Based on Berry Picking Model. Data Analysis and Knowledge Discovery, 8(4), 152-166.

[81] Bates, M. J. (1989). The design of browsing and berrypicking techniques for the online search interface. *Online review*, 13(5), 407-424.

[82] Khan, A., Hughes, J., Valentine, D., Ruis, L., Sachan, K., Radhakrishnan, A., ... & Perez, E. Debating with More Persuasive LLMs Leads to More Truthful Answers. *In Forty-first International Conference on Machine Learning*.

[83] Hua, Y., & Artzi, Y. (2024). Talk Less, Interact Better: Evaluating In-context Conversational Adaptation in Multimodal LLMs. arXiv preprint arXiv:2408.01417.

[84] Yan, L., Sha, L., Zhao, L., Li, Y., Martinez‐Maldonado, R., Chen, G., ... & Gašević, D. (2024). Practical and ethical challenges of large language models in education: A systematic scoping review. *British Journal of Educational Technology*, 55(1), 90-112.

[85] Plebe, A., & Perconti, P. (2022). *The future of the artificial mind*. Crc Press.

[86] Angwaomaodoko, E. A. (2023). The Re-examination of the Dangers and Implications of Artificial Intelligence for the Future of Scholarship and Learning. Available at SSRN 4848051.

[87] Cohen, L., Manion, L., & Morrison, K. (2002). *Research methods in education*. routledge.

[88] Li, X., Liu, M., Gao, S., & Buntine, W. (2023, August). A survey on out-of-distribution evaluation of neural NLP models. In P*roceedings of the Thirty-Second International Joint Conference on Artificial Intelligence* (pp. 6683-6691).

[89] Fagbohun, O., Iduwe, N. P., Abdullahi, M., Ifaturoti, A., & Nwanna, O. M. (2024). Beyond traditional assessment: Exploring the impact of large language models on grading practices. Journal of Artifical Intelligence and Machine Learning & Data Science, 2(1), 1-8.

[90] Cao, Y., Zhou, L., Lee, S., Cabello, L., Chen, M., & Hershcovich, D. (2023, May). Assessing Cross-Cultural Alignment between ChatGPT and Human Societies: An Empirical Study. In *Proceedings of the First Workshop on Cross-Cultural Considerations in NLP* (C3NLP) (pp. 53-67).

[91] Hardt, M., Price, E., & Srebro, N. (2016). Equality of opportunity in supervised learning. *Advances in neural information processing systems*, 29.

[92] Ji, Z., Lee, N., Frieske, R., Yu, T., Su, D., Xu, Y., ... & Fung, P. (2023). Survey of hallucination in natural language generation. *ACM Computing Surveys*, 55(12), 1-38.

[93] Barbella, M., & Tortora, G. (2022). Rouge metric evaluation for text summarization techniques. Available at SSRN 4120317.

[94] Song, F., Yu, B., Li, M., Yu, H., Huang, F., Li, Y., & Wang, H. (2024, March). Preference ranking optimization for human alignment. In *Proceedings of the AAAI Conference on Artificial Intelligence* (Vol. 38, No. 17, pp. 18990-18998).

[95] White, R. W., & Kelly, D. (2006, November). A study on the effects of personalization and task information on implicit feedback performance. *In Proceedings of the 15th ACM international conference on Information and knowledge management* (pp. 297-306).

[96] Cao, C., Sang, J., Arora, R., Kloosterman, R., Cecere, M., Gorla, J., ... & Bobrovitz, N. (2024). Prompting is all you need: LLMs for systematic review screening. medRxiv, 2024-06.

[97] Davis, J. L. (2017). Curation: A theoretical treatment. Information, *Communication & Society*, 20(5), 770-783.

[98] Stephanidis, C., Salvendy, G., Antona, M., Chen, J. Y., Dong, J., Duffy, V. G., ... & Zhou, J. (2019). Seven HCI grand challenges. *International Journal of Human‐Computer Interaction*, 35(14), 1229-1269.

[99] Rahlin, M., Barnett, J., Becker, E., & Fregosi, C. M. (2019). Development through the lens of a perception-action-cognition connection: recognizing the need for a paradigm shift in clinical reasoning. *Physical Therapy*, 99(6), 748-760.

[100] Gibson, E. J. (1988). Exploratory behavior in the development of perceiving, acting, and the acquiring of knowledge. *Annual review of psychology*, 39(1), 1-42.

[101] Little, D. Y., & Sommer, F. T. (2013). Learning and exploration in action-perception loops. *Frontiers in neural circuits,* 7, 37.

[102] Bammer, G., O'Rourke, M., O'Connell, D., Neuhauser, L., Midgley, G., Klein, J. T., ... & Richardson, G. P. (2020). Expertise in research integration and implementation for tackling complex problems: when is it needed, where can it be found and how can it be strengthened? *Palgrave Communications, 6(1)*, 1-16.

[103] Sun, M., Wang, M., & Wegerif, R. (2019). Using computer‐based cognitive mapping to improve students' divergent thinking for creativity development. B*ritish Journal of Educational Technology, 50(5)*, 2217-2233.

[104] Sim, J., Saunders, B., Waterfield, J., & Kingstone, T. (2018). Can sample size in qualitative research be determined a priori?. *International journal of social research methodology*, *21*(5), 619-634..

[105] Reeves, S. M., Crippen, K. J., & McCray, E. D. (2021). The varied experience of undergraduate students learning chemistry in virtual reality laboratories. *Computers & Education, 175*, 104320.